# Point Defects in Two-Dimensional $\gamma$-Phosphorus Carbide

*Andrey A. Kistanov[†,*], Vladimir R. Nikitenko[‡], Oleg V. Prezhdo[‡,§]*

[†]Nano and Molecular Systems Research Unit, University of Oulu, 90014 Oulu, Finland

[‡]National Research Nuclear University MEPhI, 115409 Moscow, Russia

[§]Department of Chemistry, University of Southern California, Los Angeles, CA 90089, United States

**ABSTRACT:** Defects are inevitably present in two-dimensional (2D) materials and usually govern their various properties. Here a comprehensive density functional theory-based investigation of 7 kinds of point defects in a recently produced $\gamma$ allotrope of 2D phosphorus carbide ($\gamma$-PC) is conducted. The defects, such as antisites, single C or P, and double C and P and C and C vacancies, are found to be stable in $\gamma$-PC, while the Stone−Wales defect is not presented in $\gamma$-PC due to its transition metal dichalcogenides-like structure. The formation energies, stability, and surface density of the considered defect species as well as their influence on the electronic structure of $\gamma$-PC is systematically identified. The formation of point defects in $\gamma$-PC is found to be less energetically favourable then in graphene, phosphorene, and $MoS_2$. Meanwhile, defects can significantly modulate the electronic structure of $\gamma$-PC by inducing hole/electron doping. The predicted scanning tunneling microscopy images suggest that most of the point defects are easy to distinguish from each other and that they can be easily recognized in experiments.

**TOC Graphic**

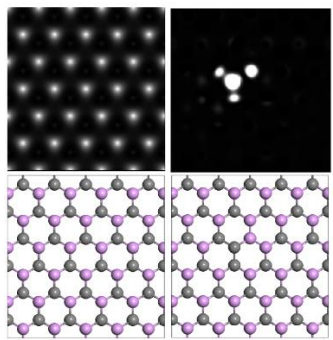

**KEYWORDS:** first-principles calculations; nanomaterials; defect engineering; phosphorus carbide; scanning tunneling microscopy

Over the past decade, two-dimensional (2D) materials have become common building blocks in various branches of science and industry.[1,2] Complex theoretical and advanced experimental studies allow obtaining new atomically thin structures, including hybrid structures of already discovered 2D materials.[3-7] Specifically, the structural similarity of graphene and phosphorene has allowed the theoretical prediction[8-10] of a new family of 2D phosphorus carbide (PC) materials consisting of carbon and phosphorus. Later, the experimental discovery[11,12] of 2D PC allotropes has been achieved by combining the theoretical predictions and previous experimental observations indicating that different ratios of $PH_3$/$CH_4$ gases during the radio frequency plasma deposition process lead to different atomic ratios of P/C, producing different structures in phosphorus carbide.[13] Depending on the geometry of their structure these 2D PC allotropes can vary from metallic to semi-metallic to wide bandgap semiconducting.[9,14,15]

One of the most extraordinary allotropes of 2D PC, $\gamma$-PC with a unique $\gamma$-InSe-like structure, has not been considered until recently. The first investigation on $\gamma$-PC has been reported by Claeyssens et al.,[16] who have considered stability and electronic structure of the bulk form of $\gamma$-PC. Later, theoretical studies have proposed an atomically thin $\gamma$-PC layer.[10,17] $\gamma$-PC can be distinguished from other PC allotropes by its superior structural stability, resistance to deformation, ultrahigh conductivity,[10] and excellent ability to recover upon interaction with environmental molecules.[17] Further, monolayer $\gamma$-PC is also predicted to show semiconducting behaviour with a direct bandgap of ~2.60 eV.[10] Despite the recent emergence, 2D allotropes of PC have already showed extremely high potential for applications in optoelectronics,[18] lithium-ion batteries,[10] single-photon source,[11] high-performance field-effect transistors,[12] and gas- and bio-sensors.[17]

It is commonly accepted that all 2D layered materials host structural defects such as point defects,[19-22] which are usually generated by ion/electron irradiation or by mechanical damage of the material.[23] Defects strongly influence material structure and properties, and therefore need to be accounted for in applications.[24-26] Similar to other 2D materials, γ-PC surfaces may contain structural defects, which may affect performance of γ-PC-based devices. However, to our knowledge, no comprehensive studies have focused on typical point defects in γ-PC up to date. Thus, the influence of point defects on the characteristics of γ-PC in applications is unknown. In this work, *ab initio* electronic structure and molecular dynamics calculations are used to systematically study the geometry, stability, and electronic properties of typical point defects in γ-PC. Theoretical guidance on the experimental routes of detection of these defects in γ-PC is provided.

Figure 1a-f shows the geometry of perfect γ-PC and the most common typical point defects found to be stable in γ-PC. These are antisite defect (AD), single vacancy of a C atom ($SV_C$), single vacancy of a P atom ($SV_P$), double vacancy of one C atom and one P atom ($DV_{CP}$), and double vacancy of two P atoms ($DV_{CC}$). It should be noted that, differently from graphene-like 2D materials, the Stone–Wales (SW) defects are not formed in transition metal dichalcogenides (TMDs) due to the polar nature of chemical bonds with trigonal symmetry.[27] Here we found that γ-PC that possesses a TMDs-like structure is also free of SW defects (Section 1 in SI and Figure S1 and movie 1). As shown in Figure 1b, AD occurs by exchanging the positions of neighbouring C and P atoms. Two kinds of SV defects can be produced by removing the C or P atom from γ-PC, as shown in Figures 1c and d, respectively. There are five kinds of divacancy (DV) defects in γ-PC. The first is the $DV_{CP}$ defect which is created when two neighbouring atoms (one C atom and one P atom) are removed from the γ-PC surface (Figure 1e). $DV_{CC}$ is formed when two C atoms are removed from the γ-PC surface (Figure 1f). The remaining DV defects formed when two C or P ($DV_{PP}$) atoms are removed from the surface or from the plane perpendicular to the surface are found to be unstable in γ-PC or leading to its destruction. More specific changes in the geometry of γ-PC induced by point defects are show in Figure S2 and are discussed in Section 2 in supporting information (SI).

The *ab initio* molecular dynamics (AIMD) simulations are performed to check the thermal stability of the considered defects in $\gamma$-PC. The initial states at the time $t = 0.0$ *ps* for the performed calculations are set to optimized geometric structures of defect-containing $\gamma$-PC. It is predicted that only AD, $SV_C$, $SV_P$, $DV_{C+P}$, and $DV_{CC}$ are stable in $\gamma$-PC at 400 K for times longer than 2.0 ps (detailed information is presented in Section 3 in SI, and Figures S3-S7 and movies 2-6). In all the remaining cases the vacancies rather annihilate, as in the case of the SW defect, or lead to the disintegration of the $\gamma$-PC structure, as in the case of the $DV_{PP}$ defect (movie 7).

The stability of the considered point defects in $\gamma$-PC is also considered in terms of their formation energy $E_{form}$, which is calculated as

$$E_{form} = E_{perfect} - E_{defect} - N_C \cdot E_C - N_P \cdot E_P \tag{1}$$

where $E_{perfect}$ and $E_{defect}$ are the total energies of perfect and defect-containing $\gamma$-PC, $E_C$ and $E_P$ are the energies of a single carbon and phosphorus atom, and $N_C$ and $N_P$ correspond to the number of the removed carbon and phosphorus atoms.

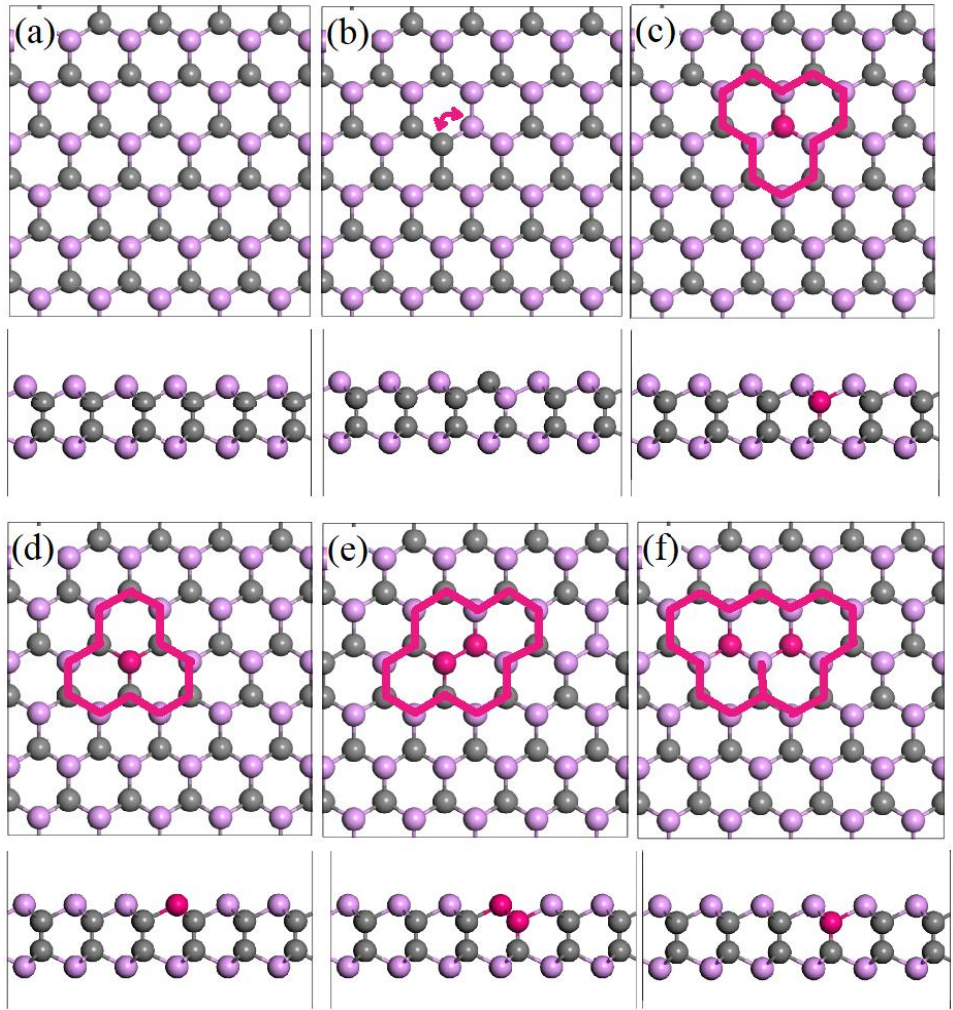

**Figure 1.** The schematic top and side views of (a) perfect and (b) AD−, (c) SV$_C$−, (d) SV$_P$−, (e) DV$_{CP}$−, and (f) DV$_{CC}$−containing γ-PC. The area of the considered point defects is marked by the pink lines. The carbon, phosphorus, and vacant atoms are colored in gray, violet, and pink.

The calculated $E_f$ of the considered stable defects in γ-PC is presented in Table 1. According to Table 1, $E_f$ of AD is the lowest among the considered defects. The SV$_P$ defect has a ~1.9 times lower $E_f$ (5.48 eV) compared to the SV$_C$ defect (10.39 eV). For comparison, $E_f$ of SV in graphene is 7.5 eV,[28] and $E_f$ of SV in phosphorene is ~1-2 eV.[29] For the most common TMDs material, MoS$_2$, $E_f$ of the SV defects varies from 2.12 (sulfur vacancy) to 6.20 eV (molybdenum vacancy).[30] Therefore, the formation of SVs in γ-PC is less favorable compared to that in phosphorene and MoS$_2$, and is similar to that in graphene. Among DVs, DV$_{CP}$ has the lowest $E_f$ = 13.80 eV, while $E_f$ of DV$_{CC}$ is much higher and is equal to 18.87 eV. For comparison, $E_f$ of DV in graphene[28] is ~8 eV, $E_f$ of DV in phosphorene[29] is ~2-3 eV, and $E_f$ of

DV in MoS$_2$[30] is ~4 eV. Thus, the formation of the DVs in γ-PC is much less favorable energetically then in graphene, phosphorene, and MoS$_2$.

**Table 1.** The calculated formation energy $E_f$ (eV) of point defects in γ-PC.

|  | AD | SV$_C$ | SV$_P$ | DV$_{CP}$ | DV$_{CC}$ |
|---|---|---|---|---|---|
| γ-PC | 1.28 | 10.39 | 5.48 | 13.80 | 18.87 |
| graphene[28] | - | 7.5 | | 8 | |
| phosphorene[29] | - | 1-2 | | 2-3 | |
| MoS$_2$[30,31] | 5.77, 5.79 | 2.12, 6.20 | | 4 | |

Further, the surface density of defects in γ-PC, at a finite temperature, is evaluated (detailed information is presented in Section 4 in SI). It is found that point defects in γ-PC have much lower surface densities compared to phosphorene[29] and MoS$_2$,[32] while the surface density of some defects such as DVs in graphene[29] is slightly higher than in γ-PC (Figure S8 in SI). This also confirms that the creation of point defects in γ-PC is less energetically favorable than in phosphorene[29] and MoS$_2$,[32] but may be more energetically favourable than in graphene.[29] The obtained results may indicate high structural stability of γ-PC, as defects usually lead to faster degradation of 2D surfaces.[29] On the other hand, the estimated comparably low thermodynamic probability of point defect formation in γ-PC can be much higher during manufacturing, as it is common for most 2D materials,[33] since synthesis may be governed by kinetic factors. Furthermore, the defects concentration in γ-PC can be controlled by means of defect engineering.[34]

The simulated scanning tunnelling microscopy (STM) images of perfect and defect-containing γ-PC are obtained to facilitate differentiation of point defects in γ-PC experimentally. Based on the expected STM characterization methods, both a constant height mode, which is faster but useful only for relatively smooth surfaces, and a constant current mode that measures irregular surfaces with high precision, but is more demanding, are considered. The STM images of the perfect γ-PC and γ-PC containing point defects are presented in Figure 2 (constant height mode) and Figure S9 (constant-current mode at a +1.0 V bias). Most defects correlate with their defective atomic structures and are easy to recognize. For instance, the AD in $\gamma$-PC can be easily recognized in both types of STM images, because it appears as one middle bright spot surrounded by smaller bright spots (Figure 2b and Figure S9b), which are contributed by the unoccupied states from the carbon and phosphorus atoms at the defect site. The STM image at the +1.0 V bias gives a clear view of the of the $SV_C$ defect with three bright spots and three grey lines connecting phosphorus atoms (Figure S9c). The $SV_P$ defect in $\gamma$-PC (Figure 2d) is also easy to recognize, it is presented by nine bright spots, which are contributed by three carbon atoms around the defect core. The $DV_{CP}$ defect (Figure 2e) show similar STM images as the $SV_C$ defect (Figure 2c) with bright spot. To distinguish between these defects in the experiments advanced methods are needed. In particular, it is possible to differentiate these two defects at the +1.0 V bias by comparing their atomic structures and STM images (Figures S9c and e). The $DV_{CC}$ defect in $\gamma$-PC can be distinguished by comparing the atomic structures and STM images (Figure 2f and Figure S9f).

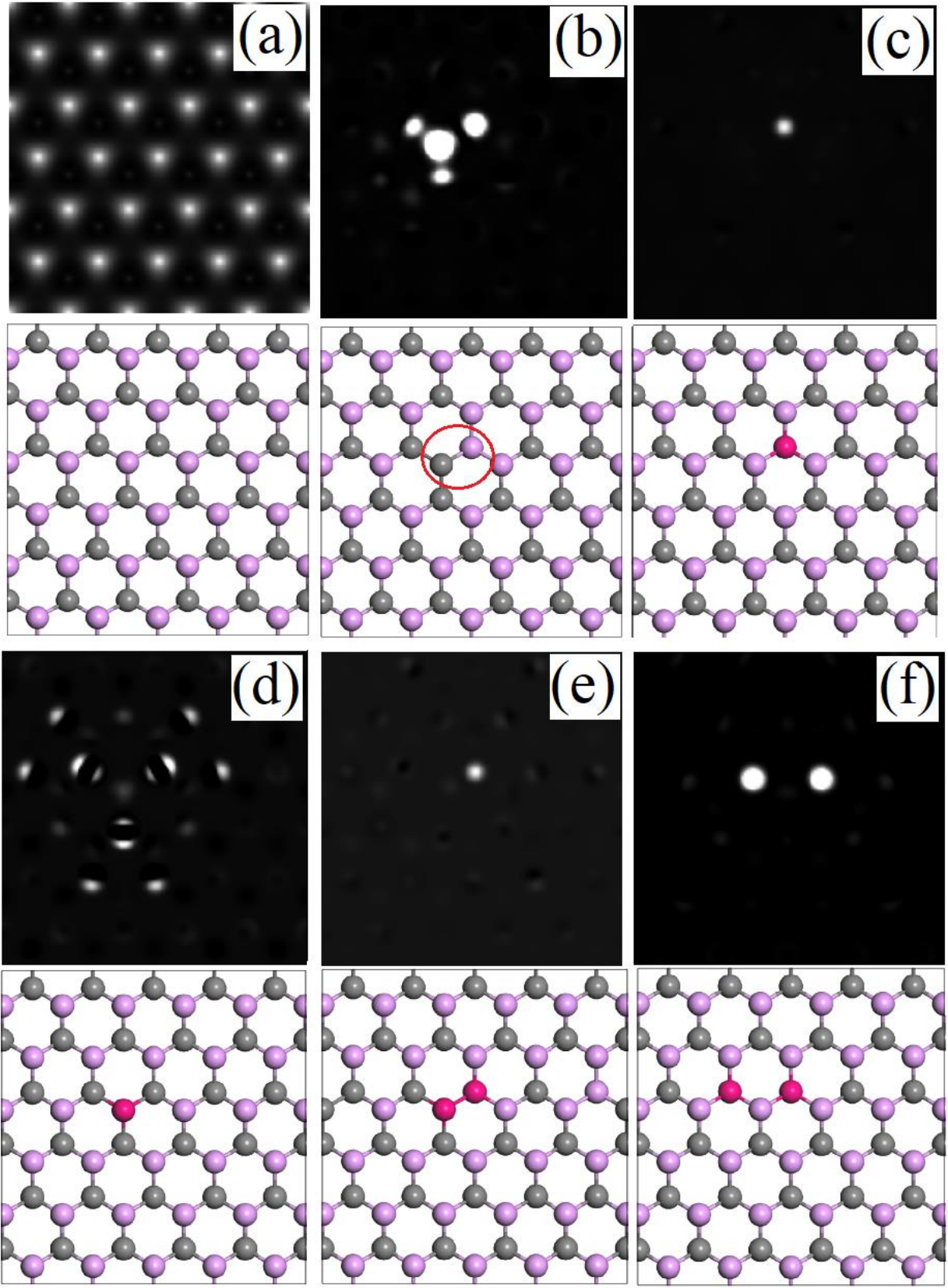

**Figure 2.** The STM images at a constant height mode (the upper panels) and the geometry (the lower panels) of (a) perfect and (b) AD−, (c) $SV_C$−, (d) $SV_P$−, (e) $DV_{CP}$−, and (f) $DV_{CC}$−containing $\gamma$-PC. The carbon, phosphorus, and vacant atoms are colored in gray, violet, and pink.

It has been proven that the local electronic properties of 2D materials may change substantially around the defect sites due to breaking of the lattice periodicity.[35-37] Figures 3a-f show the band structure plots of (a) perfect and (b) AD−, (c) SV$_C$−, (d) SV$_P$−, (e) DV$_{CP}$−, and (f) DV$_{CC}$−containing γ-PC. There is a significant realignment of bands in the band plots of vacancy-containing γ-PC, and there are non-zero states at the Fermi level in the cases of SV$_C$ and SV$_P$. These newly formed localized states in the bandgap of the host γ-PC can be attributed to the dangling bond states due to defects in γ-PC, similarly to those in phosphorene[29,38] MoS$_2$.[30,36]

Perfect γ-PC is found to be a semiconductor with an indirect bandgap of 1.51 eV according to the Perdew−Burke−Ernzerhof (PBE) exchange-correlation functional under the generalized gradient approximation (GGA) (Figure 3a), and 2.58 eV according to the hybrid exchange-correlation functional of Heyd−Scuseria−Ernzerhof (HSE) (Figure S10 in SI). This is in a good agreement with the previous studies.[10,17] Considering the lower computational load, the GGA method is used in further calculations. More detailed discussion on the bandgap calculations techniques can be found in Section 6 in SI. It should be noted that the bandgap of perfect γ-PC is larger than those of phosphorene (~0.91 eV GGA value and 1.88 eV HSE value)[29,38] and MoS$_2$ (1.8 eV HSE value).[34]

According to Figure 3b, the AD leads to a downward shift of the valence and conduction bands and appearance of defect states in the fundamental bandgap of γ-PC, 0.45 eV below the Fermi level. However, AD-containing γ-PC remains a semiconductor with almost the same indirect bandgap of 1.47 eV. Similar behavior is common for MoS$_2$ in the presence of ADs.[30] To understand the effect of concentration of ADs on the band structure of γ-PC, two ADs in are considered the same simulation cell, i.e. the AD concentration increased two times (Figure S16 in SI). Additional defect states appear below the Fermi level in the fundamental bandgap of γ-PC. On the other hand, there is no significant readjustment of bands in γ-PC which remains a semiconductor with the host bandgap of 1.47 eV. Considering the predicted low concentration of defects in γ-PC, we expect a minor influence of the defect concentration, particularly the concentration of the AD, on the electronic properties of γ-PC. For the SV$_c$ defect, removal of a carbon atom from perfect γ-PC creates unpassivated phosphorus atoms and dangling bonds in the defect core which lead to the formation of unoccupied localized states in the γ-PC fundamental bandgap, as shown in Figure 3c. A partially occupied defect-induced band, which crosses the Fermi level, appears at about 0.34 eV above the valence band maximum (VBM) of the host γ-PC. According to Figure 3d, in case of the SV$_P$ defect, the valence bands are greatly upshifted, while partially occupied defect band located at about 0.25 eV above the VBM crosses the Fermi level. This suggests easy generation of hole states in γ-PC (p-type conductivity) even upon moderate thermal excitations, which is similar to the behavior of phosphorene with a single vacancy defect.[38]

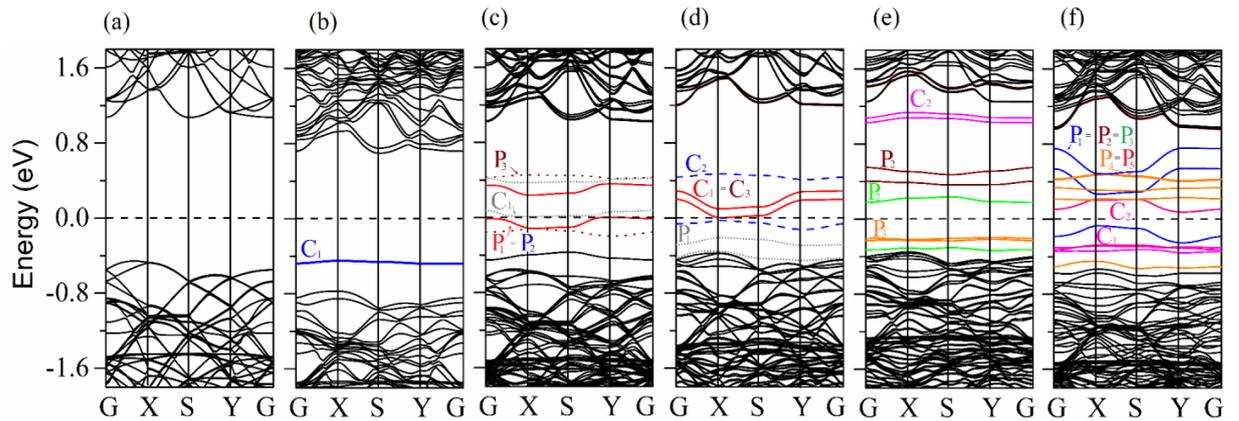

**Figure 3.** The band structure of (a) perfect and (b) AD−, (c) SV$_C$−, (d) SV$_P$−, (e) DV$_{CP}$−, and (f) DV$_{CC}$−containing γ-PC. The Fermi level is marked by the black dashed line and is set to zero. The in-gap states are labelled according to Figure 5.

Based on Figure 3e, defect states with nearly flat band dispersion for the DV$_{CP}$ defect reside inside in the fundamental bandgap of γ-PC. Although DV$_{CP}$ containing γ-PC exhibits a significant readjustment of the band lines, it still has a bandgap of 1.45 eV (0.37 eV including in-gap states). In addition, the valence and conduction bands become flatter, and as a result, the conduction band minimum (CBM) of the host γ-PC shifts from the S point (for perfect γ-PC) to the Y point, and the VBM of the host γ-PC shifts from the X point (for perfect γ-PC) to the vicinity of the Y point. This change in the band structure, signifying the possibility of an indirect-to-direct bandgap transition in $\gamma$-PC in the presence of the DV$_{CP}$ defect, may affect the optical emission efficiency of $\gamma$-PC. Importantly, as shown above, it may be hard to distinguish the SV$_C$ and DV$_{CP}$ defects in an STM image. As these defects induce remarkably different changes in the band diagram of host $\gamma$-PC, they may be distinguished from each other via photoemission spectroscopy techniques.[39] The band structure of DV$_{CC}$−containing $\gamma$-PC is shown in Figure 3f. Partially occupied defect bands are mixed with the valence bands of the host $\gamma$-PC and are located at about 0.07 eV below the Fermi level. In addition, there are defect bands located 0.07 eV above the Fermi level. Therefore, both $n$- and $p$-type conductivity is possible in $\gamma$-PC.

The density of states resolved in space known as local density of states (LDOS), computed for peripheral atoms in the defect core and for atoms far from the defect core are significantly different, which suggests that the states are renormalized greatly in the defect core. Therefore, to better understand the changes in the band structure of $\gamma$-PC due to the presence of point defects in terms of the contribution of each atom within the defect core, LDOS plots for these atoms are evaluated, as shown in Figure 4. The states introduced by the AD are mainly contributed by the $C_1$ atom and with a minor contribution from the $P_1$ atom (Figure 4a). The changes depicted at the band structure diagram of SV$_C$−containing $\gamma$-PC (Figure 3c) arise mainly from the $P_1$, $P_2$, $P_3$, and $C_1$ atoms (Figure 4b). Similarly, the defect-induced states in the band structure of SV$_P$−containing $\gamma$-PC (Figure 3d) mainly originate from the $C_1$, $C_2$, $C_3$, and $P_1$ atoms (Figure 4c). In the case of DV$_{CP}$−containing $\gamma$-PC (Figure 4d), the P1−3 (major contribution) and C1−3 atoms (minor contribution) surrounding the defect have partially occupied/unoccupied states contributing into the valence/conduction bands of the host $\gamma$-PC (Figure 3e). The P1−5 and C1−2 atoms (Figure 4e) are responsible for the in-gap states appearing at the band structure plot of the DV$_{CC}$−containing $\gamma$-PC (Figure 3f). An additional discussion of the impact of each atom in the defect core on the electronic structure of

defect−containing γ-PC is presented in Section 7 in SI, which presents projected density of states (PDOS) and provides information on the contributions of the different orbitals of the considered atoms.

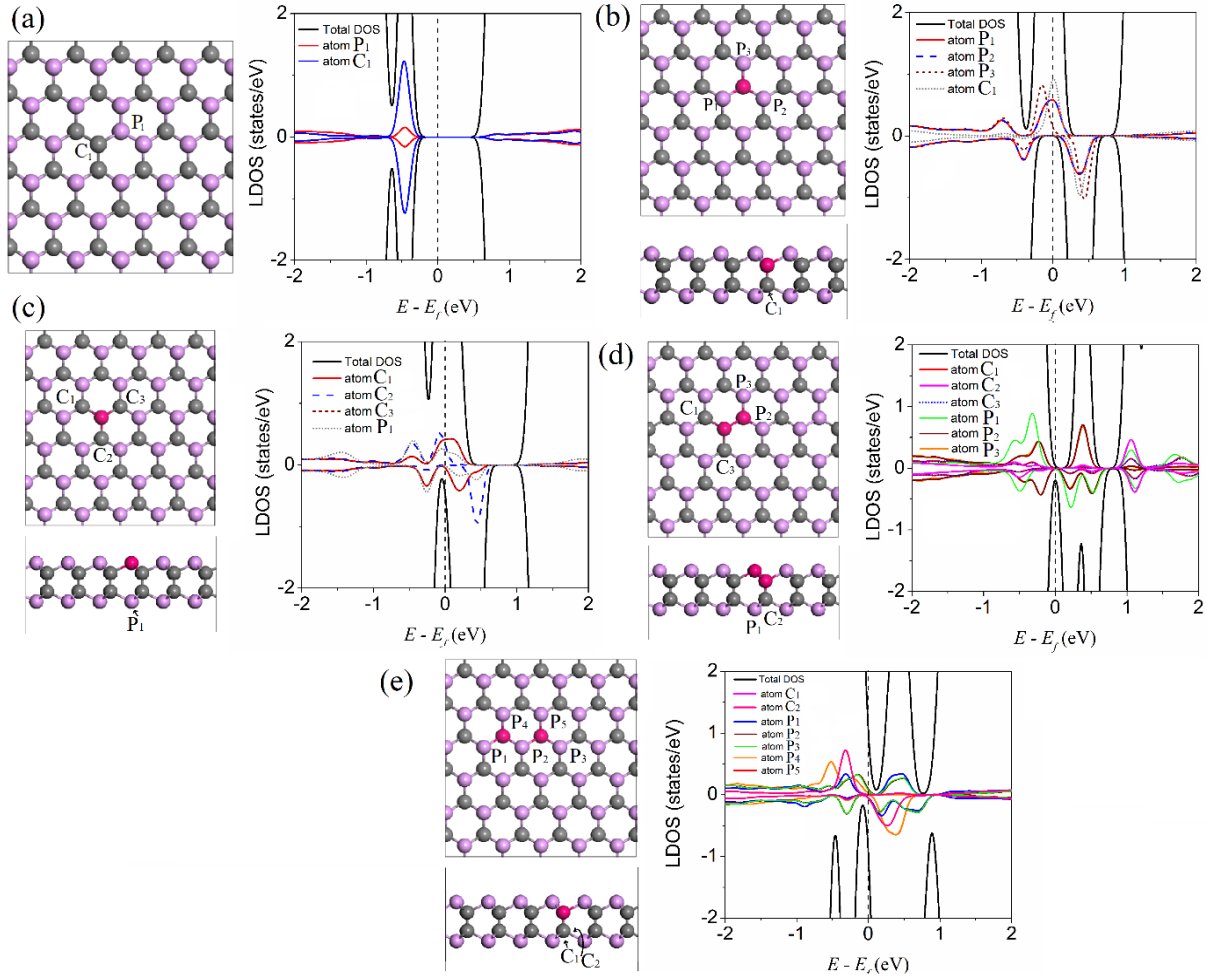

**Figure 4.** Defect geometry with atoms in the defect core labelled by numbers (left panel) and LDOS of atoms in the defect core (right panel) for (a) AD−, (b) SV$_C$−, (c) SV$_P$−, (d) DV$_{CP}$−, and (e) DV$_{CC}$−containing γ-PC. The Fermi level is marked by the black dashed line and is set to zero.

To further understand the influence of point defects on the electronic structure of γ-PC, the electron localization function (ELF) of defect-containing γ-PC is calculated. The value of the ELF maps reflects the degree of charge localization in the real space. The value ranges between 0 and 1, where 0 represents a free electronic state, while 1 represents a perfect localization. The isosurface value of 0.70 is adopted in Figure 5, in which the ELFs for AD−, SV$_C$−, SV$_P$−, DV$_{CP}$−, and DV$_{CC}$−containing γ-PC are presented. In all cases, in the area far away of the defect, the electrons are localized at the C-C and C-P bonds and P atoms, while C atoms are free of the electrons. In case of AD-containing $\gamma$-PC (Figure 5a) the electrons are clearly packed together at the $C_1$ atom, while the $P_1$ atom become empty of electrons. In addition, slight localization of electrons is observed at two P-P bonds of the $P_1$ atom in the defect core. It can be concluded that the $C_1$ atom forms strong $sp^2$ bonds with neighboring C atoms, while the $P_1$ atom has unpassivated dangling bonds. That can explain the formation of empty states within the original bandgap of $\gamma$-PC, as it is shown at the band structure plot in Figure 3b. According to Figure 5b, in case of the $SV_C$ defect, the charge simply redistributes from the removed C atom to the neighboring $P_1$, $P_2$, $P_3$, and $C_1$ atoms. As indicated in Figure 5c, the presence of the $SV_P$ defect leads to an increase of the charge localization at the $C_2$-$P_1$ bond and the $C_2$ atom in the defect core. That indicates a strong covalent chemical bonding between the $C_2$ atom in the defect core and the $P_1$ atom right below the absent P atom. Strong electron redistribution is observed in the $DV_{CP}$−containing $\gamma$-PC (Figure 5d). A full depletion of electrons from the $C_1$-$P_1$ and $C_3$-$P_1$ bonds and their almost full depletion from the $C_2$-$P_1$ bond in the defect core are found. In addition, a strong electron localization at the $P_1$ atom located right below the absent P atom and the C-C bond of the $C_1$ atom is also found. Therefore, the $C_1$-$P_1$ and $C_3$-$P_1$ bonds of the $P_1$ atom, are broken and the $C_2$-$P_1$ bond of the $P_1$ atom is nearly broken. Consequently, the presence of the $DV_{CP}$ defect leads to the formation of a hole in the $\gamma$-PC layer. As indicated in Figure 5e, the $DV_{CC}$ defect in $\gamma$-PC leads to an increase of electron localization at the $P_1$-$P_5$, $C_1$, and $C_2$ atoms and the P-P and C-P bonds of the $P_6$ atom, while there are no localized electrons at the $P_6$ atom itself. The increase in the electron localization at the bonds of the $P_6$ atom indicates the formation of a strong covalent chemical bonding between the $P_6$ and $P_2$, and $P_6$ and $C_3$ atoms.

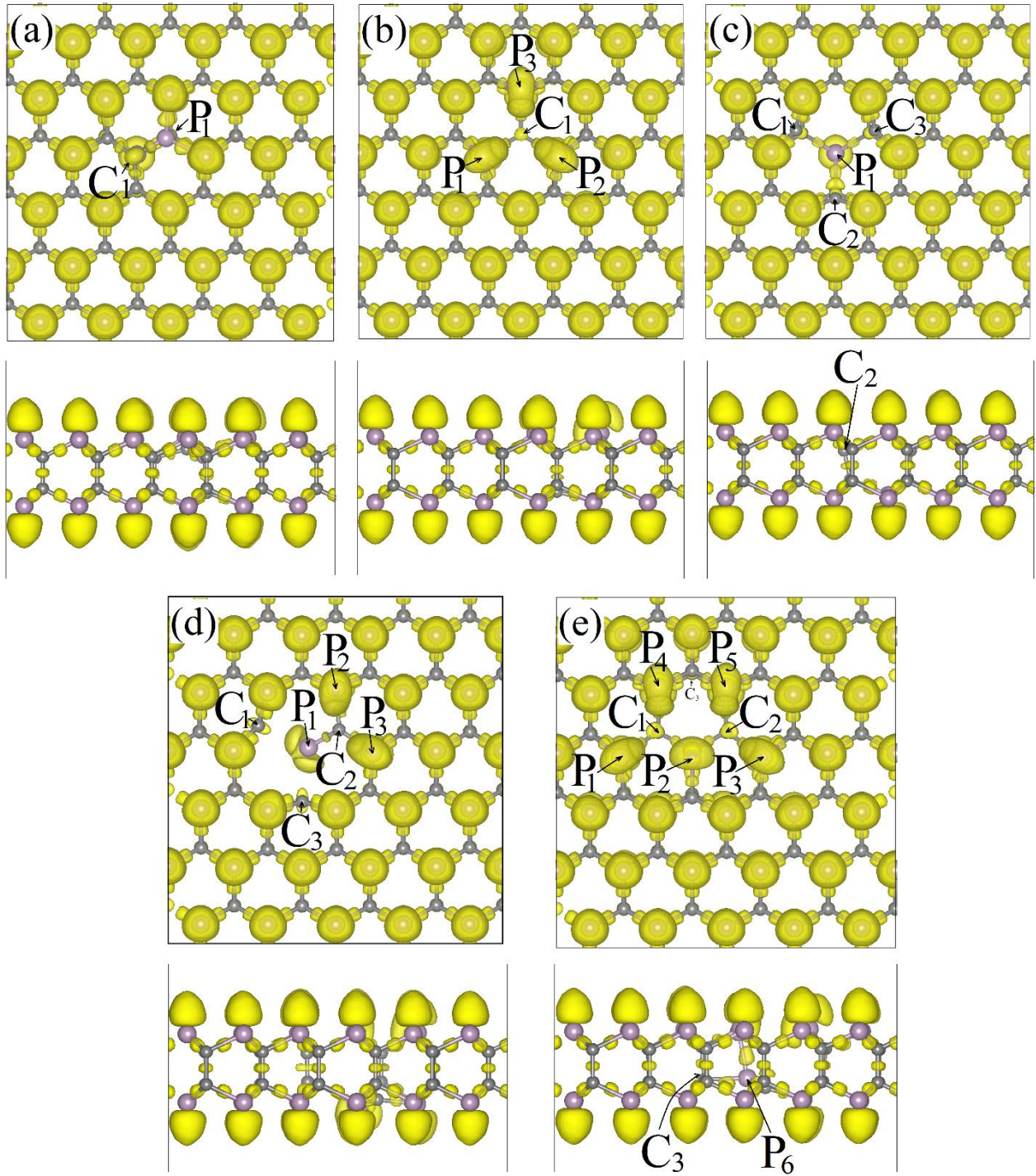

**Figure 5.** The ELFs for (a) AD−, (b) SV$_C$−, (c) SV$_P$−, (d) DV$_{CP}$−, and (e) DV$_{CC}$−containing $\gamma$-PC. The isosurface value is set to 0.7.

In summary, density-functional theory (DFT) and *ab initio* molecular dynamics calculations have been used for the systematic study of the geometry, stability, and electronic structure of typical point defects in $\gamma$-PC. It is found that the Stone-Wales defect is unlikely to form in $\gamma$-PC due to its TDMs-like structure. Nevertheless, $\gamma$-PC may host a wide variety of other point defects, such as antisites, SVs, and DVs. All these defects (accept $DV_{PP}$) are found to be highly stable in γ-PC at the temperature of 400 K. The energy input required for the formation of single and double vacancies is comparable to that for graphene and is ~2 times higher than that for phosphorene and $MoS_2$. The simulated scanning tunnelling microscopy images have shown the possibility of the identification of the considered point defects in γ-PC in experiments. The predicted stable defects also impose significant changes in the electronic structure of γ-PC, which allows their identification via photoemission spectroscopy techniques. Despite the predicted comparably low concentration of point defects in γ-PC, it may be possible to control hole/electron doping in γ-PC by defect-engineering as in graphene and phosphorene.[40] This reported work provides valuable insights into the influence of point defects on the electronic properties of γ-PC, facilitates their identification in experiments, and widens potential applications of γ-PC.

**Methods**

The spin-polarized DFT-based simulations were used in this work as implemented in the Vienna Ab initio Simulation Package.[41] The PBE GGA[42] and HSE06[43] functionals were adopted for the geometry optimization and electronic structure calculations of perfect γ-PC. In electronic structure calculations of defect-containing γ-PC the GGA PBE functional was used. The geometries of all structures were fully optimized until the total energy and all forces on the atoms converged to less than $10^{-8}$ eV and 0.01 eV/Å, respectively. The cutoff energy for plane waves was set to 400 eV. The calculated lattice parameters of γ-PC were found to be $a$ = 3.09 Å and $b$ = 5.35 Å. The considered supercell of γ-PC was composed of 5×3 unit cells (60 C and 60 P atoms). The concentration of SVs is 0.83% (one C/P-defect per 120 atoms) and the concentration of DVs is 1.67% (two C/P-defects per 120 atoms). The periodic boundary conditions were applied for the two in-plane transverse directions while the vacuum depth of 15 Å was introduced to the direction perpendicular to the surface plane to avoid artificial interactions in the considered supercells. AIMD calculations[44] were conducted at 400 K to verify the thermal stability of defects in γ-PC. The AIMD simulations lasted 2 ps or longer, and were carried out with a 1.0 fs time step. The temperature was controlled by the Nose–Hoover thermostat.

## ASSOCIATED CONTENT

The Supporting Information is available free of charge on the ACS Publications website at DOI: 10.1021/acs.jpcc.XXXX.

AIMD results on the stability of the SW−, AD−, $SV_C$−, $SV_P$−, $DV_{CP}$−, $DV_{CC}$−, and $DV_{PP}$−containing γ-PC (Figures S1, S3-S6 and Movies 1-7); the geometry of defects in γ-PC (Figure S2); the comparison of the surface density of the defects in γ-PC, graphene, phosphorene, and $MoS_2$ (Figure S8); the STM images of AD−, $SV_C$−, $SV_P$−, $DV_{CP}$−, and $DV_{CC}$−containing γ-PC (Figure S9); the comparison of the electronic structure calculations using the GGA PBE and HSE06 methods (Figure S10); the partial density of states of atoms in the defect core of defect-containing γ-PC (Figures S11-S15).

## AUTHOR INFORMATION

**Corresponding Authors**

*E-mail: andrey.kistanov@oulu.fi

**Notes**

The authors declare no competing financial interest.

ACKNOWLEDGMENTS

A.A. Kistanov acknowledges the financial support provided by the Academy of Finland (grant No. 311934). O. V. P. acknowledge funding of the U. S. National Science Foundation, grant No. CHE-1900510. The authors wish to acknowledge CSC – IT Center for Science, Finland, for computational resources.

REFERENCES

1. Liu, B.; Zhou, K. Recent progress on graphene-analogous 2D nanomaterials: Properties, modeling and applications. *Prog. Mater. Sci.* **2019**, *100*, 99−169.

2. Xiao, Z.; *et al*. Recent development in nanocarbon materials for gas sensor applications. *Sensor. Actuat. B Chem.* **2018**, *274*, 35-267.

3. Ma, D.; Zhang, J.; Lia, X.; Heb, C.; Lu, Z.; Luc, Z.; Yang, Z.; Wang, Y. $C_3N$ monolayers as promising candidates for $NO_2$ sensors. *Sens. Actuators. B*. **2018**, *266*, 664-673.

4. Lee, G. J.; Lee, M. K.; Park, J. J.; Hyeon, D. Y.; Jeong, C. K.; Park. K. I. Piezoelectric energy harvesting from two-dimensional boron nitride nanoflakes. *ACS Appl. Mater. Interfaces* **2019**, *11*(41), 37920–37926.

5. Tian, X.; Xuan, X.; Yu, M.; Mu, Y.; Lu, H. G.; Zhang, Z.; Li S. D. Predicting two-dimensional semiconducting boron carbides. *Nanoscale* **2019**, *11*, 11099-11106.

6. Kou, Z.; Guo, B.; He, D.; Zhang, J.; Mu, S. Transforming two-dimensional boron carbide into boron and chlorine dual-doped carbon nanotubes by chlorination for efficient oxygen reduction. *ACS Energy Lett.* **2018**, *3*(1), 184–190.

7. Jing, Y.; Ma, Y.; Li, Y.; Heine, T. $GeP_3$: A small indirect band gap 2D crystal with high carrier mobility and strong interlayer quantum confinement. *Nano Lett.* **2017**, *17*, 1833−1838.

8. Claeyssens, F.; Hart, J. N.; Allan, N. L.; Oliva, J. M. Solid phases of phosphorus carbide: An ab initio study. *Phys. Rev. B: Condens. Matter Mater. Phys.*, **2009**, *79*, 134115.

9. Guan, J.; Liu, D.; Zhu, Z.; Tománek, D. Two-dimensional phosphorus carbide: Competition between sp$^2$ and sp$^3$ bonding. *Nano Lett.* **2016**, *16*, 3247.

10. Zhang, W.; Yin, J.; Zhang, P.; Tang, X.; Ding, Y. Two-dimensional phosphorus carbide as a promising anode material for lithium-ion batteries. *J. Mater. Chem. A* **2018**, *6*, 12029–12037.

11. Huang, X.; *et al.* Black phosphorus carbide as a tunable anisotropic plasmonic metasurface. *ACS Photonics* **2018**, *5*, 3116–3123.

12. Tan, W. C.; *et al.* Few-layer black phosphorus carbide field-effect transistor via carbon doping. *Adv. Mater.* **2017**, *29*, 1700503.

13. Pearce, S. R. J.; May, P. W.; Wild, R. K.; Hallam, K. R.; Heard, P. J. Deposition and properties of amorphous carbon phosphide. *Diamond Relat. Mater.* **2002**, *11*, 1041–1046.

14. Shcherbinin, S. A.; Zhou, K.; Dmitriev, S. V.; Korznikova, E. A.; Davletshin, A. R.; Kistanov A. A. Two-dimensional black phosphorus carbide: rippling and formation of nanotubes. *J. Phys. Chem. C* **2020**, *124*(18), 10235–10243.

15. Wang, B. T.; Liu, P. F.; Bo, T.; Yin, W.; Eriksson, O.; Zhao, J.; Wang, F. Superconductivity in two-dimensional phosphorus carbide ($β_0$-PC). *Phys. Chem. Chem. Phys.* **2018**, *20*, 12362–12367.

16. Claeyssens, F.; Fuge, G. M.; Allan, N. L.; May, P. W.; Pearce. S. R. J.; Ashfold M. N. R. Phosphorus carbide thin films: experiment and theory. *Appl. Phys. A* **2004**, *79*, 1237–1241.

17. Kistanov, A. A. The first-principles study of the adsorption of $NH_3$, NO, and $NO_2$ gas molecules on InSe-like phosphorus carbide. *New J. Chem.* **2020**, *44*, 9377-9381.

18. Wang, G.; Pandey, R.; Karna, S. P. Carbon phosphide monolayers with superior carrier mobility. *Nanoscale* **2016**, *8*, 8819−8825.

19. Banhart, F.; Kotakoski, J.; Krasheninnikov, A. V. Structural defects in graphene. *ACS Nano* **2011**, *5*, 26−41.

20. Zhang, L.; Chu, W.; Zheng, Q.; Benderskii, A. V.; Prezhdo, O. V.; Zhao, J. Suppression of electron-hole recombination by intrinsic defects in 2D monoelemental material. *J. Phys. Chem. Lett.* **2019**, *10*(20), 6151–6158;

21. Seo, J. W. T.; Green, A. A.; Antaris, A. L.; Hersam, M. C. High-concentration aqueous dispersions of graphene using nonionic, biocompatible block copolymers. *J. Phys. Chem. Lett.* **2015**, *6*, 14, 2738–2739.

22. Zhang, L.; Vasenko, A. S.; Zhao, J.; Prezhdo, O. V. Mono-elemental properties of 2D black phosphorus ensure extended charge carrier lifetimes under oxidation: Time-domain ab initio analysis. *J. Phys. Chem. Lett.* **2019**, *10*(5), 1083–109.

23. Komsa, H. P.; Kotakoski, J.; Kurasch, S.; Lehtinen, O.; Kaiser, U.; Krasheninnikov, A. V. Two-dimensional transition metal dichalcogenides under electron irradiation: Defect production and doping. *Phys. Rev. Lett.* **2012**, *109*, 035503.

24. Liu, Y.; Xu, F.; Zhang, Z.; Penev, E. S.; Yakobson, B. I. Two-dimensional mono-elemental semiconductor with electronically inactive defects: The case of phosphorus. *Nano Lett.* **2014**, *14*, 6782−6786.

25. Hong, J. H.; *et al.* Exploring atomic defects in molybdenum disulphide monolayers. *Nat. Commun.* **2015**, *6*, 6293.

26. Cai, Y.; Ke, Q.; Zhang, G.; Yakobson, B. I.; Zhang, Y. W. Highly itinerant atomic vacancies in phosphorene. *J. Am. Chem. Soc.* **2016**, *138*(32), 10199–10206.

27. Wu, Z.; Ni, Z. Spectroscopic investigation of defects in two-dimensional materials. Nanophotonics **2017,** *6*(6), 1219–1237.

28. Krasheninnikov, A. V.; Lehtinen, P. O.; Foster, A. S.; Nieminen, R. M. Bending the rules: Contrasting vacancy energetics and migration in graphite and carbon nanotubes. *Chem. Phys. Lett.* **2006**, *418*, 132–136.

29. Hu, W.; Yang, J. Defects in phosphorene. *J. Phys. Chem. C* **2015**, *119*, 20474−20480.

30. Hong, J.; et al. Exploring atomic defects in molybdenum disulphide monolayers. *Nat Commun.* **2015**, *6*, 6293.

31. Wang, P.; Qiao, L.; Xu, J.; Li, W.; Liu, W. Erosion mechanism of $MoS_2$-based films exposed to atomic oxygen environments. *ACS Appl. Mater. Interfaces* **2015**, *7*(23), 12943–12950.

32. Jeong, H. Y.; *et al.* Visualizing point defects in transition-metal dichalcogenides using optical microscopy. *ACS Nano* **2016**, *10*, 770−777.

33. Cai, Y.; Ke, Q.; Zhang, G.; Zhang, Y. W. Energetics, charge transfer and magnetism of small molecules physisorbed on phosphorene. *J. Phys. Chem. C* **2015**, 119, 3102–3110.

34. Hu, Z.; Wu, Z.; Han, C.; He, J.; Ni, Z.; Chen, W. Two-dimensional transition metal dichalcogenides: interface and defect engineering. *Chem. Soc. Rev.* **2018**, 47(9), 3100–3128.

35. Zhou, Q.; Chen, Q.; Tong, Y.; Wang, J. Light-induced ambient degradation of few-layer black phosphorus: Mechanism and protection. *Angew. Chem., Int. Ed.* **2016**, 128, 11609–11613.

36. Cai, Y.; Zhou, H.; Zhang, G.; Zhang, Y. W. Modulating carrier density and transport properties of $MoS_2$ by organic molecular doping and defect engineering. *Chem. Mater.*, **2016**, 28, 8611-8621.

37. Kistanov, A. A.; Cai, Y.; Zhou, K.; Srikanth, N.; Dmitriev, S. V.; Zhang, Y. W. Exploring the charge localization and band gap opening of borophene: a first-principles study. *Nanoscale* **2018**, *10*, 1403-1410.

38. Kistanov, A. A.; Cai, Y.; Zhou, K.; Dmitriev, S. V.; Zhang, Y. W. The role of $H_2O$ and $O_2$ molecules and phosphorus vacancies in the structure instability of phosphorene. *2D Mater.* **2017**, *4*, 015010.

39. Hamer, M. J.; *et al.* Indirect to direct gap crossover in two-dimensional InSe revealed by angle-resolved photoemission spectroscopy. *ACS Nano* **2019**, *13*(2), 2136–2142.

40. Hersam, M. C. Defects at the two-dimensional limit. *J. Phys. Chem. Lett.* **2015**, 6(14), 2738–2739.

41. Kresse, G.; Furthmüller, J. Efficient iterative schemes for ab initio total-energy calculations using a plane-wave basis set. *Phys. Rev. B: Condens. Matter Mater. Phys.* **1996**, *54*, 11169.

42. Perdew, J. P.; Burke, K.; Ernzerhof, M. Generalized gradient approximation made simple. *Phys. Rev. Lett.* **1996**, *77*, 3865.

43. Heyd, J.; Scuseria, G. E.; Ernzerhof, M. Hybrid functionals based on a screened Coulomb potential. *J. Chem. Phys.* **2003**, *118*, 8207.

44. Henkelman, G.; Uberuaga, B. P.; Jonsson, H. A climbing image nudged elastic band method for finding saddle points and minimum energy paths. *J. Chem. Phys. 2000*, *113*, 9901.